# Transport-defined photodetection through electrically selectable nonequilibrium carrier transport

Jiaxuan Cai[1,2], Yan Xie[1,2], Ning Yang[3,4], Weidong Chu[3,4], Peng Bai[3,4,*], Lianhe Li[5], Edmund H. Linfield[5], Hanbing Wang[6], Meng Chen[1,2], Guanchen Li[1,2], Yingxin Wang[1,2,*], Ziran Zhao[1,2,*]

1 Department of Engineering Physics, Tsinghua University, Beijing 100084, China

2 Key Laboratory of Particle and Radiation Imaging of Ministry of Education, Tsinghua University, Beijing 100084, China

3 Institute of Applied Physics and Computational Mathematics, Beijing 100088, China

4 National Key Laboratory of Computational Physics, Beijing 100088, China

5 School of Electronic and Electrical Engineering, University of Leeds, Leeds LS2 9JT, UK

6 Microsystem & Terahertz Research Center, China Academy of Engineering Physics, Chengdu 610200, China

*e-mail: bai_peng@iapcm.ac.cn; wangyingxin@tsinghua.edu.cn; zhaozr@mail.tsinghua.edu.cn

## Abstract

Broadband optical fields can change in both intensity and spectral distribution, but a fixed-response detector maps this evolving information onto a single electrical signal. Here we demonstrate transport-defined photodetection, in which electrical bias selects how photoexcited carriers are redistributed, escape and are collected, creating complementary response functions within one shared active region. In a GaAs/AlGaAs semiconductor ratchet, light-driven ratchet transport defines the response at 0 V, whereas the spectral evolution at −2 V is consistent with field-assisted hot-carrier transport. The overlapping states span a measured spectral range of 0.4–94.5 μm at 5 K, provide state-dependent calculated detection floors and support joint infrared operation at 30 K. Direct optical beat notes at 7.432 GHz in the mid-infrared and 17.103 GHz in the terahertz demonstrate optical-to-electrical conversion. Their common-path outputs recover an imposed spatial temperature gradient and the transient field-of-view-integrated effective radiation temperature of laser-excited graphite. These results establish post-photoexcitation transport as a function-defining design variable for semiconductor photodetectors, complementing structure-defined, field-tuned and optically encoded approaches to reconfigurable photodetection.

Broadband optical fields in ultrafast spectroscopy, laser–matter diagnostics and transient process monitoring evolve jointly in intensity, spectral distribution and time[1-3]. Resolving these coupled spectral–temporal signatures can help track rapid changes in energy release, transport and thermal state in complex nonequilibrium systems, including fusion plasmas and combustion environments[4-6]. Yet a fixed-response detector integrates this information into a single current. As a result, physically distinct optical states can produce indistinguishable electrical outputs even when the detector is sensitive and fast[7,8]. Spectrometers and multichannel systems recover more information, but dispersive optics, scanning and detector arrays complicate compact common-path measurements of rapidly changing fields[9,10].

Compact multicolour and multispectral detectors address this problem by mapping incident radiation onto several electrical outputs[8,11,12]. They are effective when the relevant signatures fall within predefined spectral windows, but the number and weighting of those outputs are commonly set during material growth or fabrication[13-18]. Discrete channels can therefore undersample continuous broadband redistribution, while added channels increase matching, calibration and readout complexity[8,9,11]. This motivates mechanisms that generate multiple electrical observables from one broadband field without multiplying optical paths or fixing every spectral weighting in advance[10,19-25].

Carrier transport already determines detector current through the sequence of photoexcitation, redistribution, escape and collection[12], but it has rarely been engineered as the selectable variable that defines the measurement function. Ratchet structures provide such an opportunity: asymmetric potentials make nonequilibrium escape direction- and energy-dependent, while an applied field reshapes injection and escape[26]. Earlier GaAs/AlGaAs ratchet studies used this band profile to extend response into the far-infrared and terahertz and demonstrated suppressed dark current, operation at elevated cryogenic temperatures and fast photoelectric conversion[27-29]. The same field sensitivity points to a less-explored strategy distinct from selecting which photons are absorbed: electrically selecting how the carriers created after broadband photoexcitation are redistributed and collected.

Here we demonstrate transport-defined photodetection in a high-speed semiconductor ratchet detector (HSRD), in which bias selects two nonequilibrium transport regimes within one GaAs/AlGaAs active region. Light-driven ratchet transport at 0 V defines a low-dark-current Ratchet-defined state with a 27.5 μm operational cut-off, whereas spectral evolution consistent with field-assisted hot-carrier transport at −2 V defines a Field-assisted state extending to 94.5 μm. Together, the states provide measured spectral coverage from 0.4 to 94.5 μm at 5 K, support joint infrared operation at 30 K and retain direct optical-to-electrical conversion demonstrated by beat notes at 7.432 GHz in the mid-infrared and 17.103 GHz in the terahertz; normalized microwave rectification gives a 37.16 GHz 3-dB roll-off at −3 V. Their common-path outputs recover spatial and transient thermoradiometric information that a single fixed broadband current cannot disambiguate. The work therefore establishes ratchet-enabled transport selection as a route from intensity-only detection towards compact measurements of coupled spectral and temporal dynamics.

## Results

### Transport-defined spectral interrogation in a common active region

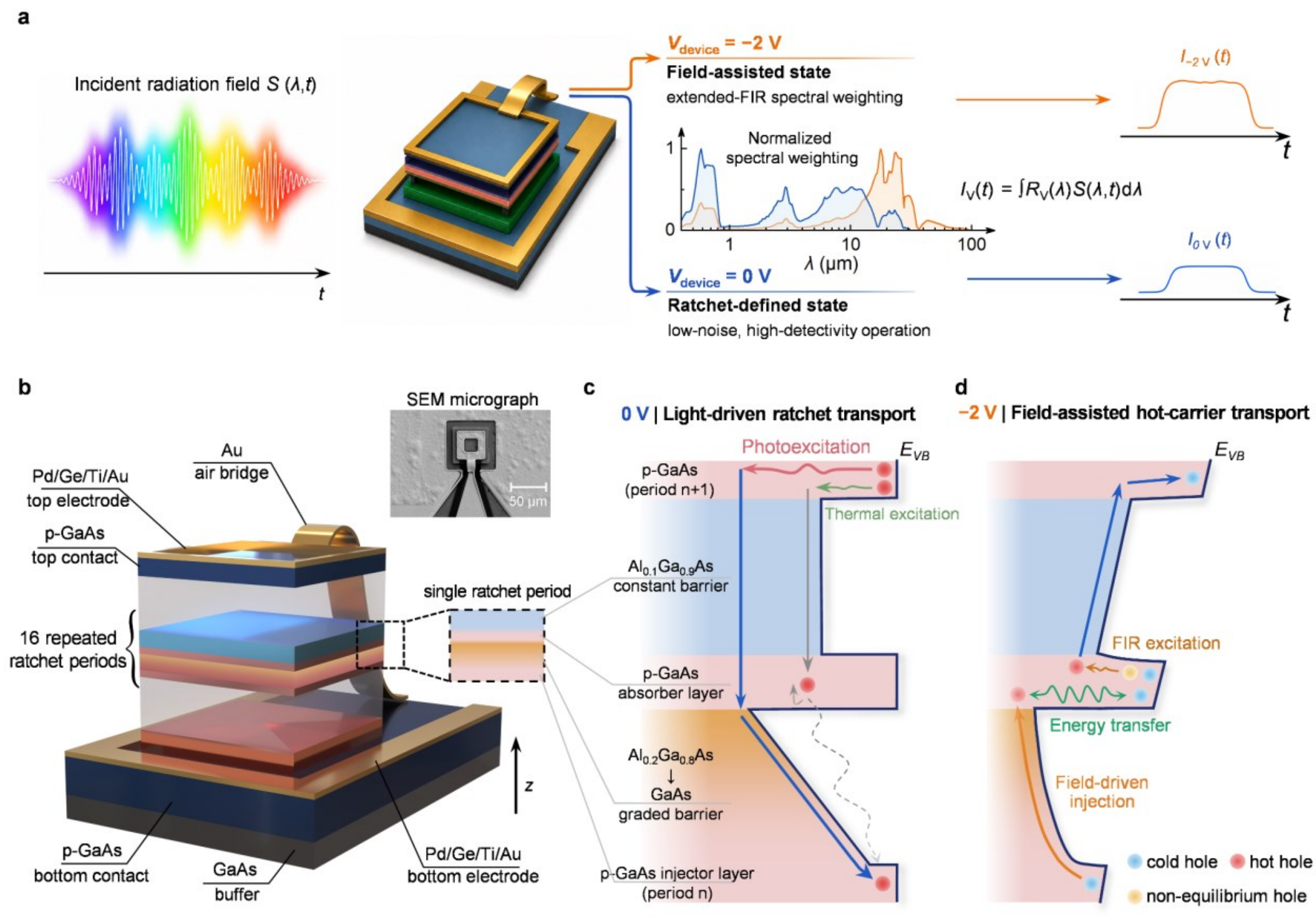

**Figure 1. Electrically selectable transport-defined measurement functions. a,** A time-varying broadband field, S($\lambda$,t), is interrogated at 0 and −2 V. The Ratchet-defined and Field-assisted states apply distinct normalized spectral weightings, $R_V(\lambda)$, to generate bias-encoded currents $I_V$(t) = ∫$R_V(\lambda)$S($\lambda$,t)d$\lambda$. The 0 V state has the lower calculated detection floor (Figure 4). **b,** HSRD architecture with 16 repeated p-type GaAs/AlGaAs ratchet periods in a 50 μm × 50 μm mesa, Pd/Ge/Ti/Au semiconductor ohmic contacts and an Au air bridge for coplanar-waveguide extraction. Insets show the fabricated device and one ratchet period; the graded barrier is $Al_xGa_{1-x}As$ with $x$ increasing from 0 to 0.20. **c,** At 0 V, inequivalent escape pathways rectify photoexcited-hole motion, producing light-driven ratchet transport while limiting thermally activated leakage. **d,** Schematic interpretation of the Field-assisted state at −2 V, in which field-driven injection and energy redistribution establish a nonequilibrium hole population, and subsequent far-infrared excitation enables barrier crossing and collection.

In the HSRD, electrical bias does not select separate absorbers or optical paths; it selects how a broadband photoexcited carrier population is redistributed, escapes and is collected within one shared active region (Figure 1a). The resulting transport regimes apply different spectral weightings to the same incident field through a common device and extraction path.

The HSRD comprises 16 repeated p-type GaAs/AlGaAs ratchet periods processed into a 50 μm × 50 μm mesa (Figure 1b). Each period contains an $Al_xGa_{1-x}As$ graded barrier, with $x$ increasing from 0 to 0.20 along the growth direction, two p-GaAs active layers and an $Al_{0.1}Ga_{0.9}As$ constant barrier. Pd/Ge/Ti/Au top and bottom ohmic electrodes contact the p-GaAs layers, and an Au air bridge connects the top electrode to the coplanar-waveguide signal line. Unless otherwise stated, the measurements in Figures 2–5 use this same device.

The active region provides the common broadband photoexcitation basis. In established interpretations of p-type GaAs/AlGaAs ratchet photodetectors, interband processes contribute in the visible and near-infrared, whereas valence-band and free-carrier processes access progressively lower photon energies in the mid- and far-infrared[27-29]. These processes generate carriers over a broad energy range. Bias then selects how those carriers are redistributed, escape through the repeated potential landscape and enter the external circuit. The detector states are therefore defined by post-photoexcitation transport rather than by switching between distinct absorbers or changing the intrinsic bandgap.

At 0 V, optical excitation drives the hole population away from equilibrium (Figure 1c). The constant and graded barriers provide inequivalent escape pathways, so photoexcited holes leave adjacent ratchet periods with different probabilities in opposite directions. This imbalance rectifies their motion and produces light-driven ratchet transport without an applied field[26,27]. Carrier confinement and the absence of field-driven injection suppress the dark current, defining the low-dark-current Ratchet-defined state.

At −2 V, the applied field tilts the repeated potential and changes the injection and escape conditions (Figure 1d). Together with the measured spectral evolution and prior hot-carrier interpretations in related GaAs/AlGaAs ratchet structures[27,28], this behaviour is consistent with field-assisted hot-carrier transport. In this picture, field-driven injection and energy redistribution increase the contribution of lower-energy excitations to barrier crossing and collection. The 0 and −2 V conditions therefore represent two carrier-selection regimes within the same architecture; Figure 2 tests whether they produce stable broadband response functions with distinguishable spectral weighting.

## Electrical selection of transport-defined broadband response functions

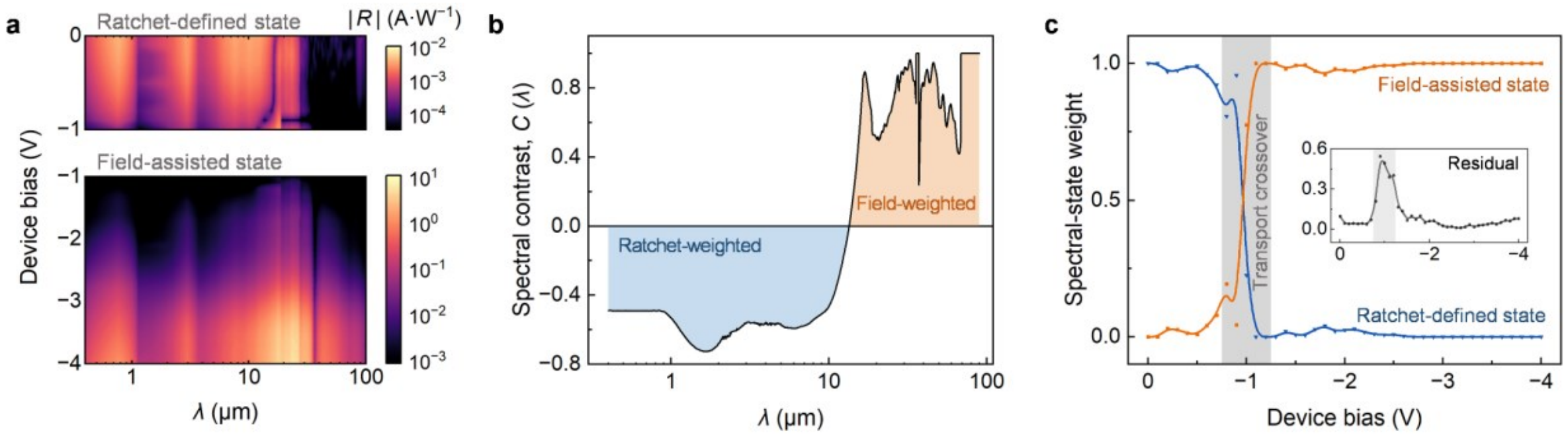

**Figure 2. Electrical selection of two broadband transport-defined states. a,** Bias- and wavelength-dependent responsivity magnitude, |R|, measured at 5 K from 0 to −4 V. Separate logarithmic colour scales display the Ratchet-defined and Field-assisted regimes. The representative 0 and −2 V states have operational long-wavelength cut-offs of 27.5 and 94.5 µm, respectively, and together span 0.4–94.5 µm. **b,** Spectral contrast between the independently normalized 0 and −2 V responsivity magnitudes. Negative and positive values denote greater relative weighting by the Ratchet-defined and Field-assisted states, respectively; both responses remain broadband and strongly overlap. **c,** Bias-dependent spectral-state weights obtained by non-negative decomposition of the normalized signed responsivity spectra using the exact measured 0 and −2 V spectra as the Ratchet-defined and Field-assisted reference functions. The weights are descriptive spectral-shape coefficients rather than carrier fractions. The shaded region marks the low-response crossover; the inset shows the normalized spectral-shape residual.

To test whether bias selects two stable broadband response functions within the common active region, rather than continuously reshaping a single response spectrum, we examine the responsivity magnitude measured from 0 to −4 V at 5 K (Figure 2a). The map separates into two regimes divided by a low-response crossover near −1 V. Within each regime, bias mainly changes the response magnitude, whereas the normalized spectral shape changes most strongly across the crossover. We therefore use the measured 0 and −2 V spectra as representative Ratchet-defined and Field-assisted response functions, respectively. Their operational long-wavelength cut-offs are 27.5 and 94.5 µm, and together they span 0.4–94.5 µm.

A small negative bias tilts the potential against the spontaneous ratchet contribution, suppressing the net response near −1 V. Beyond the crossover, the response recovers with a different spectral weighting, and the measured evolution is consistent with field-assisted hot-carrier transport. The −2 V spectrum retains shorter-wavelength sensitivity while increasing the relative contribution of lower-energy excitations. More-negative biases predominantly scale the response magnitude while the normalized spectral shape remains comparatively stable, identifying the −2 V spectrum as a representative Field-assisted response function.

The wavelength-dependent complementarity of the representative response functions is quantified by the spectral contrast $C(\lambda) = [\hat{R}_{-2\,\mathrm{V}}(\lambda) - \hat{R}_{0\,\mathrm{V}}(\lambda)]/[\hat{R}_{-2\,\mathrm{V}}(\lambda) + \hat{R}_{0\,\mathrm{V}}(\lambda)]$, where the 0 and −2 V responsivity magnitudes are independently normalized (Figure 2b). Negative contrast denotes greater relative weighting by the Ratchet-defined response function, whereas positive contrast denotes greater weighting by the Field-assisted response function. Because this normalization removes the absolute response amplitude, $C(\lambda)$ isolates differences in relative spectral shape. The two functions remain strongly overlapping while emphasizing different parts of the same broadband field, thereby providing a differential observable unavailable from either output alone.

A non-negative two-function decomposition describes how the measured bias evolution is organized by the exact 0 and −2 V reference functions (Figure 2c). The Ratchet-defined weight dominates near 0 V and the Field-assisted weight beyond the crossover. The residual remains low within the stable regimes and rises near −1 V, where the response is weak and the normalized spectral shape changes rapidly, identifying the crossover as the principal region of spectral evolution between the two response functions. Complementary 60 K TCAD analysis provides further support for bias-induced tilting of the ratchet potential and local band-edge evolution. We next test whether these response functions retain fast optical-to-electrical conversion.

**High-speed photoresponse retained across transport-defined states**

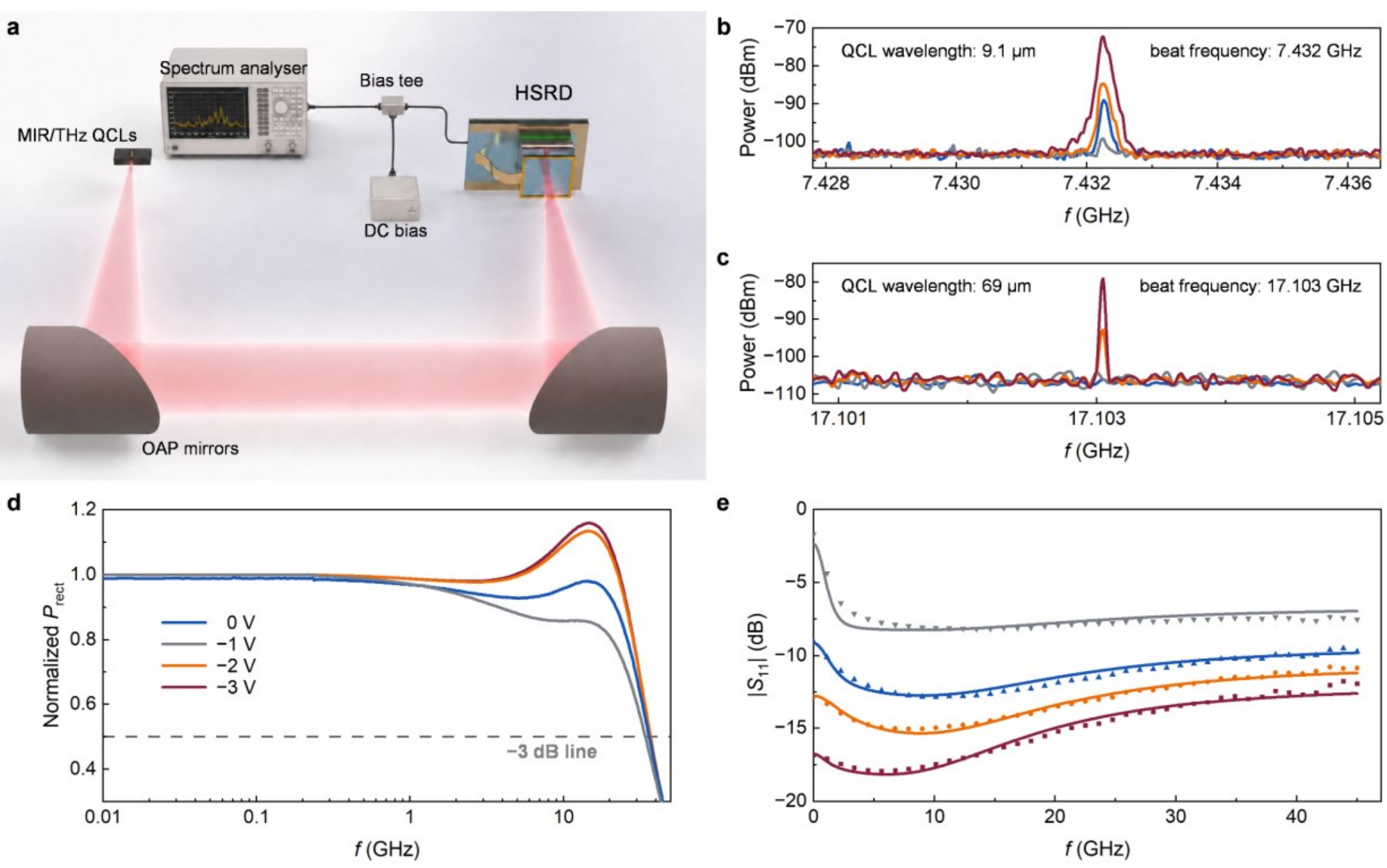

**Figure 3. High-frequency optoelectronic and radiofrequency response. a,** Optical beat-note setup using mid-infrared or terahertz quantum cascade lasers, two off-axis parabolic mirrors, the HSRD, a bias tee and a spectrum analyser. **b,** Mid-infrared beat-note spectra for a QCL wavelength of 9.1 μm, directly demonstrating optical-to-electrical response at 7.432 GHz across the bias evolution. **c,** Terahertz beat-note spectra for a QCL wavelength of 69 μm, directly demonstrating response at 17.103 GHz in the Field-assisted regime. **d,** Normalized microwave-rectification power at 0, −1, −2 and −3 V. The dashed line marks the 3-dB power level; the maximum sampled roll-off is 37.16 GHz at −3 V. **e,** Measured reflection magnitude $|S_{11}|$ (symbols) and phenomenological equivalent-circuit fits (solid lines), describing bias-dependent radiofrequency coupling and parasitic extraction.

Fast spectral interrogation also requires optical-to-electrical conversion compatible with the selected transport weightings. We distinguish direct optical beat notes from the normalized microwave-rectification envelope, while $|S_{11}|$ provides separate information on bias-dependent impedance and radiofrequency extraction.

At 9.1 µm, the HSRD resolves a 7.432 GHz beat note across the bias evolution: it is present in the Ratchet-defined state, suppressed near the crossover and recovered under stronger negative bias (Figure 3b). At 69 µm, the Field-assisted regime directly resolves a 17.103 GHz beat note (Figure 3c). These are selected-frequency demonstrations of mid-infrared and terahertz optical-to-electrical conversion.

The normalized microwave-rectification power remains nearly flat at low frequency and has a maximum sampled 3-dB roll-off of 37.16 GHz at −3 V (Figure 3d). Under high-frequency electrical excitation, it characterizes the nonlinear device and its radiofrequency injection and extraction path. Together with the beat notes, it provides an experimentally grounded reference for assessing photoelectric-conversion bandwidth and practical state-addressing rates[30,31]. Bias-dependent $|S_{11}|$ spectra and phenomenological circuit fits show operating-point-dependent coupling and parasitics (Figure 3e), consistent with field-dependent extraction and transport modifying small-signal impedance and high-frequency response.

Thus, the Ratchet-defined state retains the demonstrated gigahertz mid-infrared conversion, while the Field-assisted regime supports the demonstrated mid-infrared and terahertz frequencies.

## Operating windows for bias-encoded spectral interrogation

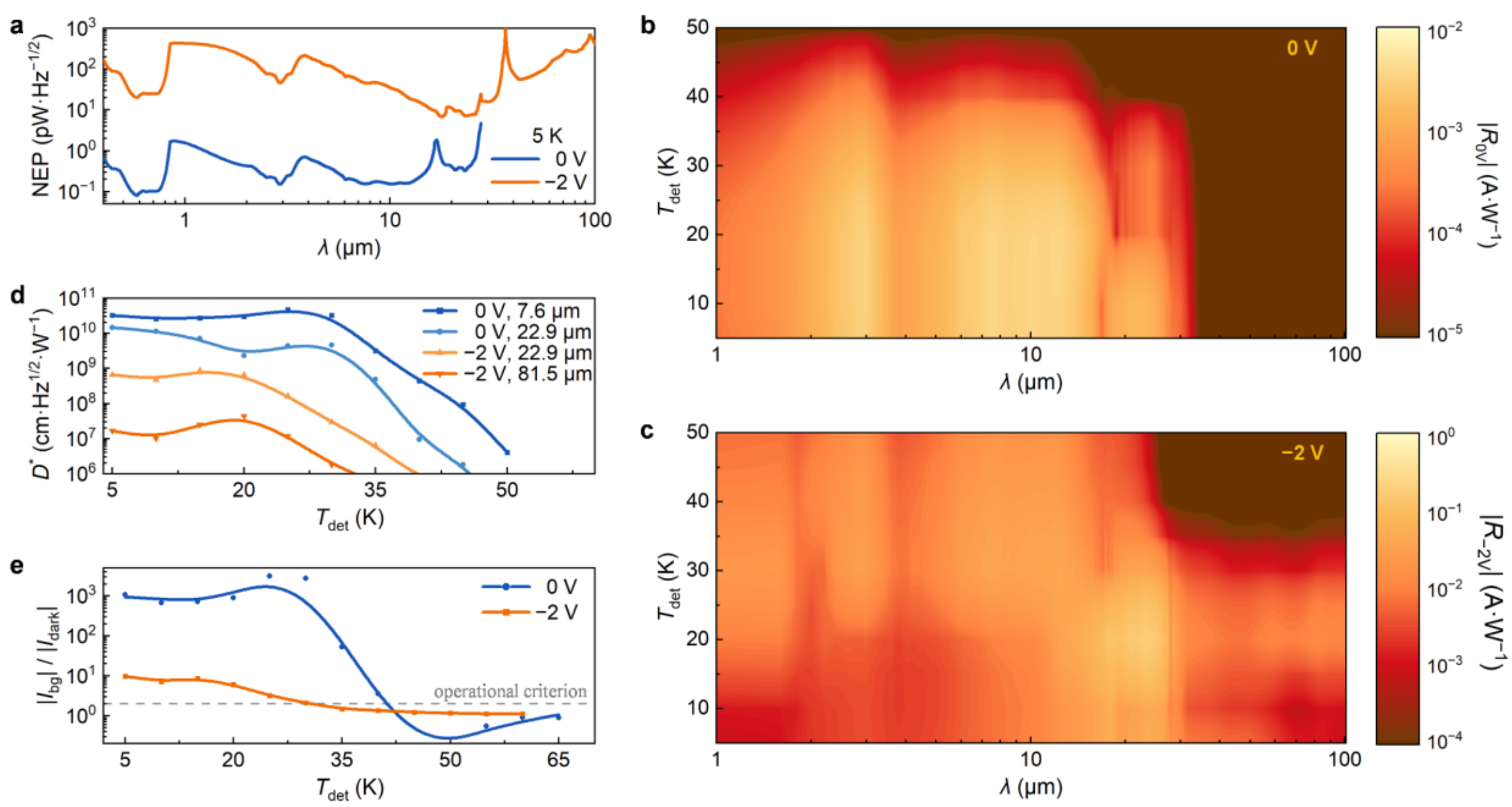

**Figure 4. Calculated detection floor, temperature response and joint operating window. a,** Calculated noise-equivalent power (NEP) at 5 K under 0 and −2 V, obtained from measured dark current and calibrated responsivity under the adopted unit-gain generation–recombination-noise model. **b,c,** Temperature-dependent magnitude of the infrared responsivity at b, 0 V and c, −2 V, where $T_{det}$ is the detector temperature. Independent logarithmic colour scales are used for the two maps. The 0 V response remains comparatively stable at low temperature and weakens above approximately 35 K. At −2 V, far-infrared and terahertz response remains observable at 30 K, while the longest-wavelength extent contracts from its 5 K value and shorter-wavelength components persist to higher temperatures. **d,** Calculated specific detectivity, $D^*$, at selected wavelengths and biases. **e,** Magnitude ratio of the total current under room-temperature background radiation, $|I_{bg}|$, to the dark current, $|I_{dark}|$. The dashed line marks the adopted operational criterion $|I_{bg}| = 2|I_{dark}|$. Together, the temperature-dependent spectral-response, calculated-detectivity and current-ratio data establish a joint usable infrared operating window verified through 30 K.

We next evaluate the detection floor and temperature stability that determine whether the two response functions can serve as complementary measurement projections.

At 5 K, the Ratchet-defined state has a substantially lower calculated NEP across its infrared range because zero-bias operation suppresses dark current (Figure 4a). Extending the cut-off from 27.5 to 94.5 μm in the Field-assisted state incurs a higher calculated NEP, especially at the longest wavelengths. The states therefore provide a lower-floor reference projection and a long-wavelength-weighted projection at higher noise cost.

With increasing detector temperature, the 0 V responsivity remains comparatively stable at low temperature and weakens above approximately 35 K (Figure 4b–d). At −2 V, far-infrared and terahertz response remains observable at 30 K, although the longest-wavelength extent contracts as thermally activated carriers obscure the lowest-energy contribution; shorter-wavelength components persist to higher temperatures. Calculated $D^*$ likewise distinguishes the stronger mid- and long-wave infrared detectivity of the Ratchet-defined state from the narrower temperature window of the Field-assisted far-infrared contribution.

Both states approach the operational criterion $|I_{bg}| = 2|I_{dark}|$ over approximately 30–35 K (Figure 4e). Together, the temperature-dependent spectra, calculated detectivity and current ratios establish a joint usable infrared operating window verified through 30 K. At 5 K, these states provide the complementary projections used in Figure 5.

## Bias-encoded interrogation of thermal radiation

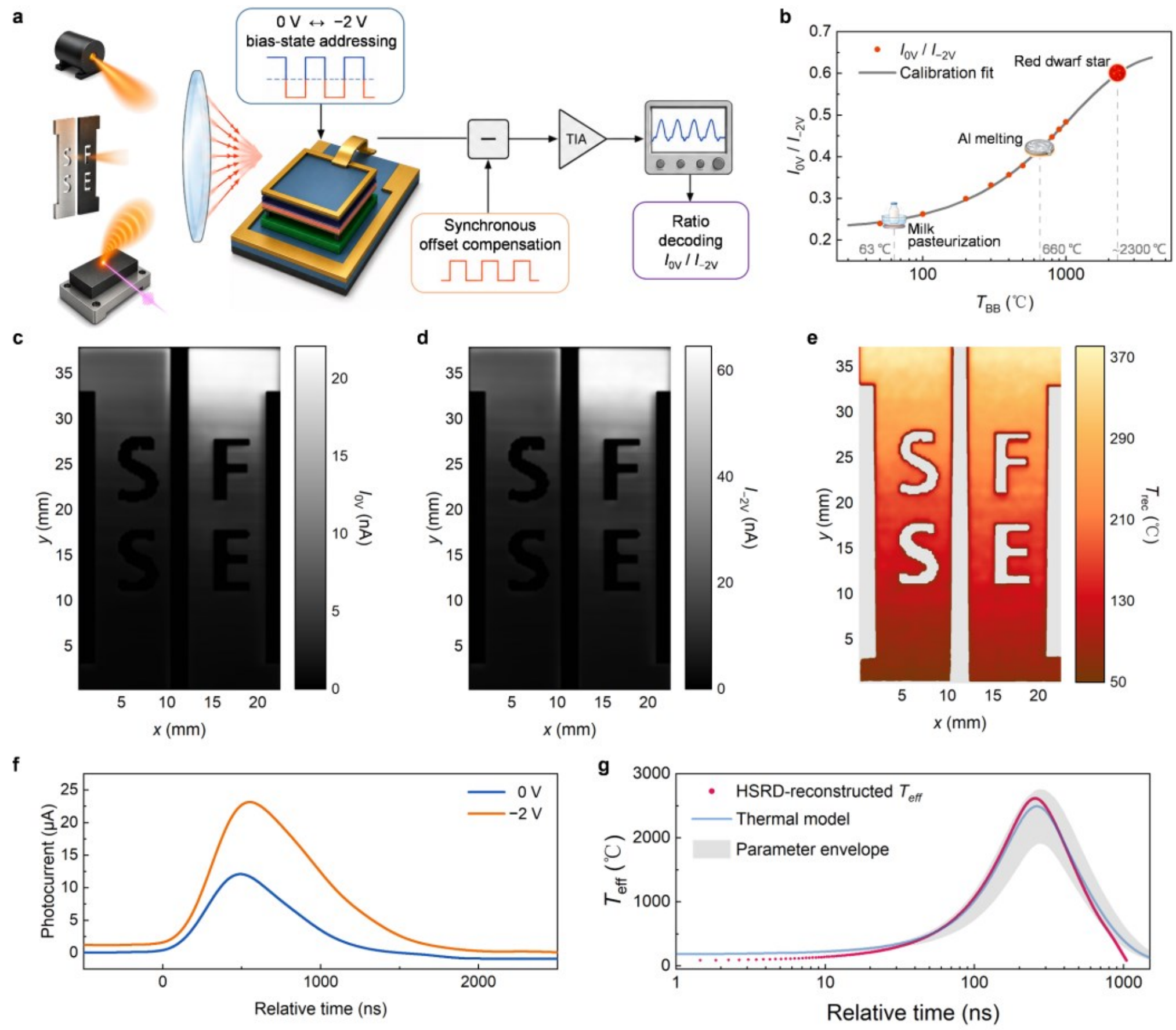

**Figure 5. Bias-encoded thermoradiometric information recovery. a,** Three scenes use the same HSRD optical readout: blackbody radiation for calibration, steady radiation from a dual-material target for spatial reconstruction and femtosecond-laser-induced transient radiation from graphite. A 1 MHz waveform demonstrates 0/−2 V addressing and synchronous offset compensation; blackbody calibration and spatial scanning use 1 kHz operation in the SR570-based chain, whereas transient traces are acquired separately at fixed biases. These rates reflect the implemented measurement conditions. **b,** Semi-empirical calibration of $I_{0\,\mathrm{V}}/I_{-2\,\mathrm{V}}$ against blackbody temperature $T_{\mathrm{BB}}$, constrained by measurements from 50 to 1,000 °C and extrapolated for the high-temperature transient reconstruction. Contextual markers indicate milk pasteurization at 63 °C[32], aluminium melting at 660 °C[33] and the approximately 2,300 °C effective temperature of TRAPPIST-1[34]. **c,d,** Dark-subtracted net photocurrent maps at c, 0 V and d, −2 V for a target combining sandblasted stainless steel and blackened grade-45 carbon steel under a common vertical thermal gradient. The materials and surface treatments provide contrasting radiometric conditions; thermocouples near the heated and cooled contact regions provide boundary references of approximately 388 and 40 °C. **e,** Hard-edge finite-support temperature reconstruction, $T_{\mathrm{rec}}$, from the calibrated current ratio; non-sample regions are masked. **f,** Baseline-subtracted transient photocurrents generated by approximately 100 fs, 800 nm pulses at 1 kHz and a nominal sample pulse energy of 50 μJ. Broadband thermal emission from graphite is recorded separately at 0 and −2 V on a relative time axis. **g,** Discrete HSRD-reconstructed field-of-view-integrated effective radiation temperatures, $T_{\mathrm{eff}}$, on the relative time axis. The blue curve is a parameter-constrained reduced-order thermal-diffusion estimate; the grey region is the pointwise 10th–90th percentile sensitivity to the selected model parameters.

Figure 5 tests the central information advantage of transport-defined photodetection: whether two complementary broadband projections measured through one optical path can recover spatial and transient information that remains underdetermined from either fixed response alone.

A 1 MHz waveform demonstrates 0/−2 V addressing with synchronous offset compensation, whereas blackbody calibration and raster scanning use 1 kHz in the SR570 readout chain (Figure 5a). These are implemented measurement rates; Figure 3 provides a separate small-signal reference. The dark-subtracted ratio $I_{0\,\mathrm{V}}/I_{-2\,\mathrm{V}}$ varies monotonically over the measured 50–1,000 °C range and is represented by a semi-empirical fit (Figure 5b); only the high-temperature transient reconstruction uses its extrapolated continuation. The reference markers illustrate the progressive shift of thermal-radiation weighting towards shorter wavelengths with increasing temperature.

Under a common imposed vertical thermal gradient, the 0 and −2 V maps of the dual-material target contain different mixtures of temperature and emissivity information (Figure 5c,d). The calibrated ratio combines the two projections, partially suppresses the material-dependent amplitude differences present in either map alone and recovers the imposed gradient in the hard-edge finite-support reconstruction (Figure 5e). This is an information-recovery result rather than a claim of emissivity-independent thermometry.

Ultrafast 800 nm excitation produces localized graphite heating and broadband thermal emission recorded separately at fixed biases of 0 and −2 V (Figure 5f). The baseline-subtracted traces share a laser-synchronous trigger and acquisition grid. Their pointwise ratio yields a field-of-view-integrated effective radiation temperature, $T_{\mathrm{eff}}$, weighted by collection geometry, graphite emissivity and the two response functions.

$T_{\mathrm{eff}}$ reaches approximately $2.6 \times 10^3$ °C at a relative time of approximately 255 ns and then decays on sub-microsecond to microsecond timescales (Figure 5g). The peak uses the extrapolated calibration and the time axis remains relative. A parameter-constrained thermal-diffusion estimate follows the trend; the grey band denotes pointwise 10th–90th percentile sensitivity to selected parameters. Thus, transport-selected common-path projections recover both spatial gradients and transient effective radiation temperatures.

## Mechanistic origins of reconfigurable detector states

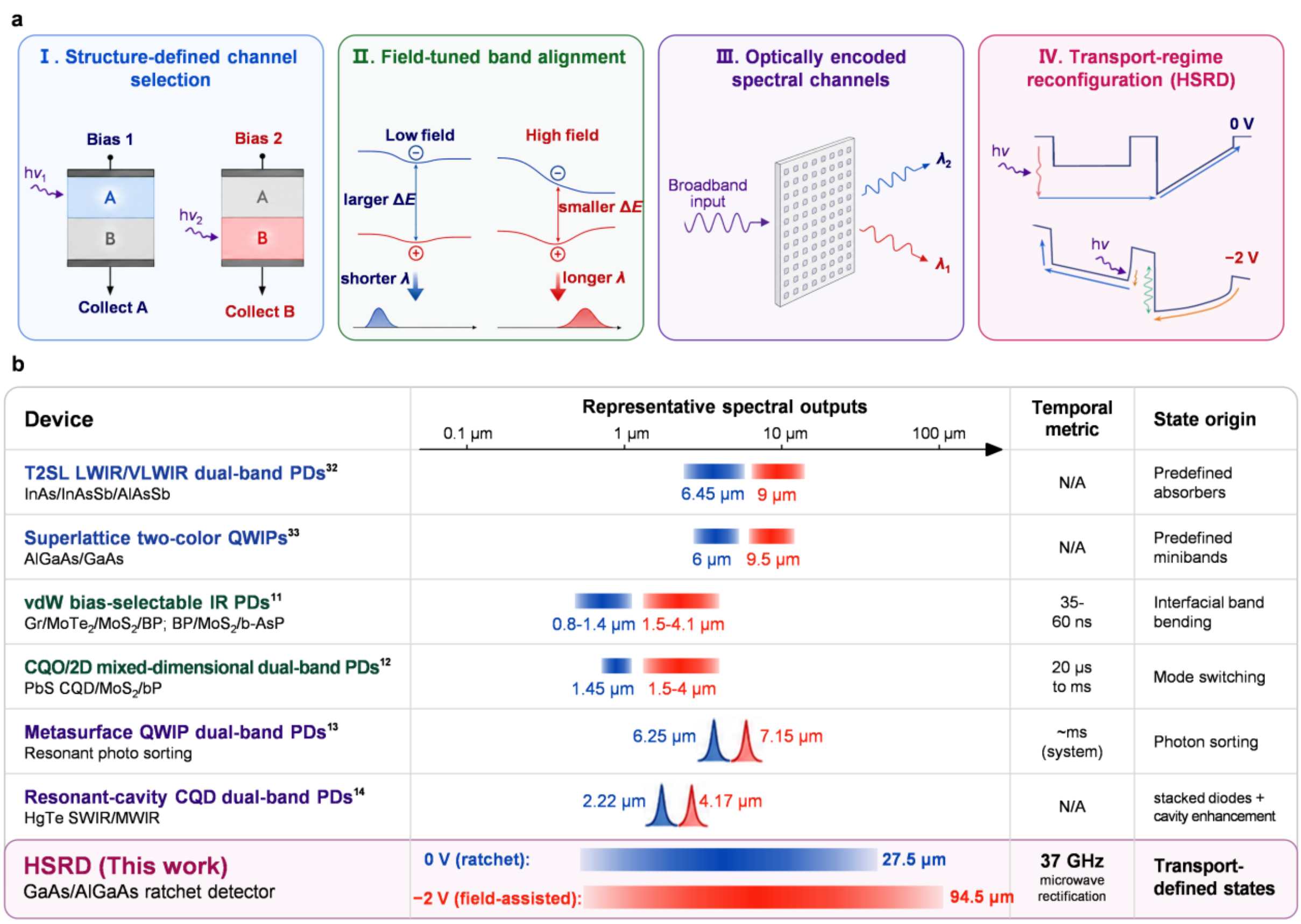

| Device | Representative spectral outputs (0.1 μm – 1 μm – 10 μm – 100 μm) | Temporal metric | State origin |
|---|---|---|---|
| T2SL LWIR/VLWIR dual-band PDs[32] InAs/InAsSb/AlAsSb | 6.45 μm 9 μm | N/A | Predefined absorbers |
| Superlattice two-color QWIPs[33] AlGaAs/GaAs | 6 μm 9.5 μm | N/A | Predefined minibands |
| vdW bias-selectable IR PDs[11] Gr/MoTe$_2$/MoS$_2$/BP; BP/MoS$_2$/b-AsP | 0.8-1.4 μm 1.5-4.1 μm | 35-60 ns | Interfacial band bending |
| CQO/2D mixed-dimensional dual-band PDs[12] PbS CQD/MoS$_2$/bP | 1.45 μm 1.5-4 μm | 20 μs to ms | Mode switching |
| Metasurface QWIP dual-band PDs[13] Resonant photo sorting | 6.25 μm 7.15 μm | ~ms (system) | Photon sorting |
| Resonant-cavity CQD dual-band PDs[14] HgTe SWIR/MWIR | 2.22 μm 4.17 μm | N/A | stacked diodes + cavity enhancement |
| HSRD (This work) GaAs/AlGaAs ratchet detector | 0 V (ratchet): 27.5 μm; −2 V (field-assisted): 94.5 μm | 37 GHz microwave rectification | Transport-defined states |

**Figure 6. Reconfigurable photodetection paradigms and representative comparison. a,** Four routes to multiple electrical spectral outputs: I, structure-defined channel selection; II, field-tuned band alignment; III, optically encoded spectral channels; and IV, transport-regime reconfiguration in the HSRD. **b,** Representative spectral outputs, temporal metrics and state origins for selected reconfigurable and dual-band photodetectors. Literature rows show reported peak wavelengths or response ranges, whereas the HSRD row shows the operational cut-offs of its two overlapping broadband states. The HSRD row therefore pairs these operational cut-offs with a 37 GHz microwave-rectification metric. Reported temporal metrics originate from different measurement methods and are not strictly equivalent; the HSRD 7.432 and 17.103 GHz beat notes are the direct optical demonstrations. The examples provide mechanistic context rather than a comprehensive performance ranking.

Figure 6 classifies representative reconfigurable and dual-band photodetectors by the dominant origin of their multiple spectral outputs. Structure-defined devices select absorbers, junctions or minibands fixed during fabrication; field-tuned devices alter band alignment, extraction conditions or device mode; optically encoded devices use resonances, cavities or photon sorting[14-17,35,36]. The categories identify dominant mechanisms rather than mutually exclusive classes.

The HSRD instead uses a ratchet structure to select post-photoexcitation redistribution, escape and collection. Light-driven ratchet transport defines the 0 V state, whereas the −2 V spectral evolution is consistent with field-assisted hot-carrier transport. Although bias tilts the ratchet potential, the response functions are not attributed to switching predefined absorbers, selecting optical modes or reconfiguring an intrinsic gap. Structural engineering instead enables the carrier-transport rule itself to become state-defining.

Panel b supplies condition-specific experimental context, not a performance ranking. The reported peak wavelengths, ranges and temporal metrics are not directly equivalent to the HSRD operational cut-offs or microwave-rectification result; the 7.432 and 17.103 GHz beat notes remain the direct optical demonstrations.

## Discussion

These results show that the measurement function applied to a broadband optical field can be defined after photoexcitation through carrier transport. Stable spectral shapes within the two bias ranges, the zero-bias ratchet response and TCAD-supported electrostatic evolution support two regimes in one active region; the negative-bias evolution remains consistent with field-assisted hot-carrier transport. Earlier studies established broadband photovoltaic ratchet response and bias-dependent spectral extension[27-29]. The advance here is to use ratchet-enabled transport selection not only to influence performance, but to create distinct detector response functions.

Such functions require co-design of the transport landscape and electrical readout. Confinement stabilizes the low-dark-current Ratchet-defined state but can restrict long-wavelength escape; field-assisted injection and redistribution increase low-energy weighting while raising dark current and narrowing the far-infrared temperature window. Broadband response, direct conversion at the demonstrated gigahertz frequencies and state-dependent calculated detection floors show complementary utility, although spectral weighting, noise and speed cannot be optimized independently. The objective is a stable set of projections rather than one maximized metric.

Two overlapping response functions provide a second broadband observable without separating the optical path or active region. Because they weight the same field differently, they can distinguish spectral redistribution that may produce similar total signals in a fixed-response detector. In Figure 5, combining the spatial projections partially suppresses material-dependent amplitude contrast and recovers the imposed gradient, while separately acquired transient projections yield a field-of-view-integrated effective radiation temperature. This is not complete spectral reconstruction; it is a demonstration that calibrated transport-defined projections can recover otherwise underdetermined spatial and transient information.

The two states remain jointly usable at 30 K, where the −2 V state retains far-infrared and terahertz response but with a contracted longest-wavelength extent; usability remains wavelength- and noise-dependent. The spatial ratio does not eliminate wavelength-dependent emissivity[37]. Transient $T_{\mathrm{eff}}$ relies on calibration extrapolation, a relative time axis and separately acquired traces; real-time switching and simultaneous acquisition were not demonstrated. The model band represents parameter sensitivity, not a confidence interval or full uncertainty budget.

Optical conversion and electrical state addressing involve distinct timescales. The demonstrated 1 MHz addressing and 1 kHz measurement rates were set by the SR570-based chain, whereas measured picosecond-scale intersubband relaxation in related p-doped

GaAs/AlGaAs quantum wells[38] and the 37.16 GHz microwave-rectification roll-off provide separate electronic-response references. Practical addressing will also depend on device charging, impedance matching, the CPW, contacts and packaging; the microwave envelope is a small-signal reference, not a full-swing switching bandwidth. Faster drivers and differential readout could enable time-interleaved acquisition, although fixed-band, room-temperature or high-resolution tasks may favour other architectures.

More broadly, transport-defined photodetection suggests co-engineering post-photoexcitation redistribution, relaxation, escape and collection with optical absorption. Such calibrated projections could help track rapid changes in energy release, transport and thermal state in nonequilibrium systems whose broadband radiation evolves jointly in intensity, spectral distribution and time. A common photoexcitation region could support multiple calibrated response functions if the transport regimes are distinct, stable and reproducible. This is not arbitrary spectral programmability; it treats transport as a function-defining variable matched to the information content and timescale of a measurement.

Electrical bias therefore selects two post-photoexcitation transport regimes within one shared GaAs/AlGaAs ratchet active region: light-driven ratchet transport at 0 V and a −2 V spectral evolution consistent with field-assisted hot-carrier transport. Their overlapping response functions combine complementary spectral weighting and calculated detection floors with broadband coverage, direct gigahertz optical-to-electrical conversion and joint infrared operation at 30 K. The common-path thermoradiometric experiments show why this distinction matters: the two projections recover spatial and transient information that a single fixed broadband current cannot determine. Ratchet-enabled transport selection thus establishes post-photoexcitation transport as a function-defining design variable for semiconductor photodetectors, complementary to structure-defined, field-tuned and optically encoded routes to reconfigurable photodetection.

## Methods

**Device growth and fabrication.** The heterostructure was grown by molecular-beam epitaxy on a semi-insulating GaAs substrate. From substrate to surface, it comprised a 700 nm Be-doped p-GaAs bottom contact/buffer, 16 repeated ratchet periods and an 80 nm Be-doped p-GaAs top contact; all p-GaAs contact and active layers were nominally doped to $1 \times 10^{19}$ cm$^{-3}$. Each period contained an 80 nm $Al_xGa_{1-x}As$ graded barrier ($x = 0 \rightarrow 0.20$ along the growth direction), a 20 nm p-GaAs layer, an 80 nm $Al_{0.1}Ga_{0.9}As$ constant barrier and a second 20 nm p-GaAs layer. A 50 µm × 50 µm mesa was defined by photolithography and wet chemical etching to a nominal depth of 3.5 µm. Pd/Ge/Ti/Au (25/75/30/200 nm) formed the semiconductor ohmic contacts; a Ti/Au seed layer with electroplated Au formed the coplanar waveguide and a nominally 10 µm-wide suspended air bridge.

**Spectral measurements and spectral-state analysis.** The HSRD was mounted in a Janis ST-100 liquid-helium cryostat, with photocurrent recorded by an SR570 current preamplifier. Visible–near-infrared responsivity was calibrated against monochromator power measured after the quartz cryostat window along the common optical path. Mid- and far-infrared FTIR response shapes were corrected for the corresponding cryostat-window transmission and relative source-plus-beamsplitter response, then placed on an absolute scale using bias-resolved net photocurrents from an IR-564 cavity blackbody. A signed broadband responsivity matrix in A $W^{-1}$ was retained for analysis. Operational cut-offs were located with a third-order Savitzky–Golay filter (51-point window) used only for boundary detection: the first local minimum within the terminal 15% of each valid long-wavelength interval was assigned as the cut-off. For spectral contrast, the exact measured 0 and −2 V responsivity magnitudes were normalized independently, with points below 3% of the maximum summed normalized response masked only for this calculation, not for cut-off extraction. Bias-dependent signed spectra from 1–100 μm were decomposed by log-wavelength-weighted non-negative least squares using the exact measured 0 and −2 V spectra as references; the coefficients are descriptive spectral-shape weights, not carrier populations or absolute responsivity amplitudes. The 60 K exact-device TCAD calculation describes static electrostatic band-edge evolution only and does not simulate nonequilibrium carrier populations, optical absorption or responsivity.

**High-frequency characterization.** Optical beat-note signals from multimode quantum cascade lasers at 9.1 and 69 μm were relayed to the HSRD by two off-axis parabolic mirrors and recorded electrically with a spectrum analyser while the device was biased through a bias tee. Microwave rectification under d.c. biases of 0, −1, −2 and −3 V used high-frequency electrical excitation; each power trace was normalized to its low-frequency value, and 37.16 GHz was the sampled frequency closest to the 0.5 power level at −3 V without interpolation or frequency-response fitting. Reflection magnitude $|S_{11}|$ was measured with a vector network analyser and fitted phenomenologically using four series-connected modules representing the CPW, top contact, active region and bottom contact. The active-region module comprised a bias-dependent series resistance followed by a parallel resistance/capacitance branch; fitted parameters describe bias-dependent impedance and parasitic coupling and were not used to infer an intrinsic RC speed limit.

**Electrical, noise and temperature characterization.** A Keithley 2410 source-measure unit was used for electrical biasing and for measurements of dark current and room-temperature background current. NEP and $D^*$ were calculated from measured dark current and calibrated responsivity using a unit-gain generation–recombination-noise model. For $g = 1$ and $\Delta f =$ 1 Hz, $i_n = \sqrt{(4q|I_{dark}|)}$, NEP = $i_n/|R|$ and $D^* = \sqrt{A}/\text{NEP}$ (equivalently $D^* = |R|/\sqrt{(4q|J_{dark}|)}$ in current-density form). The model excludes additional Johnson, 1/f, background-photon and readout-noise terms; dark current and responsivity are measured inputs, whereas noise current, NEP and $D^*$ are calculated/model-based quantities.

**Bias-encoded thermoradiometry.** All measurements used the HSRD at 5 K with a diamond cryostat window. A 1 MHz square-wave experiment demonstrated 0/−2 V bias-state addressing with synchronous electrical-offset suppression. Blackbody calibration and dual-material raster scanning used 1 kHz addressing, an SR570 transimpedance gain of $1 \times 10^6$ V $A^{-1}$ and bias-specific dark subtraction. The current ratio was fitted over measured blackbody temperatures of 50–1,000 °C as $I_{0\,V}/I_{-2\,V} = 0.64838 - 0.42444\exp(-T_{BB}/1070.9)$, with $T_{BB}$ expressed numerically in °C, and inverted pixel-wise; continuation above 1,000 °C is extrapolation used only for the high-temperature transient reconstruction. Quantitative spatial reconstruction used the hard-edge finite-support procedure; no soft-edge result was used. For transient measurements, approximately 100 fs, 800 nm pulses at 1 kHz with a nominal sample pulse energy of 50 μJ excited graphite. The 0 and −2 V traces were acquired separately on the same laser-synchronous time base with a FEMTO HSA-Y-1-60 amplifier ($5 \times 10^4$ V $A^{-1}$; 1.1 GHz), and their ratio was converted to $T_{eff}$ on a relative time axis. A reduced-order thermal-diffusion model treated the graphite sheet as an effective anisotropic medium and represented observation, diffusion and deposited-energy scales rather than a full spatial–spectral forward model. Eight parameters were sampled independently and uniformly for 3,000 trajectories, each aligned at its first rising-edge crossing of the 150 °C model reference. The displayed grey band is the pointwise 10th–90th percentile envelope of all 3,000 aligned trajectories and is a parameter-sensitivity envelope rather than a statistical confidence interval.

## Competing interests

The authors declare no competing interests.

## Use of generative AI

During the preparation of this manuscript, the authors used ChatGPT to support manuscript checking and improve language clarity. The authors reviewed and edited the manuscript as needed and take full responsibility for its content.